\documentclass[conference,10pt]{IEEEtran}
\IEEEoverridecommandlockouts
\usepackage{cite}
\usepackage{amsmath,amssymb,amsfonts}
\usepackage{algorithmic}
\usepackage{graphicx}
\usepackage{textcomp}
\usepackage{xcolor}
\def\BibTeX{{\rm B\kern-.05em{\sc i\kern-.025em b}\kern-.08em
    T\kern-.1667em\lower.7ex\hbox{E}\kern-.125emX}}

\usepackage{amsxtra,mathrsfs,dsfont,booktabs}
\usepackage[capitalize]{cleveref}

\graphicspath{{../figures/}}

\usepackage{todonotes}
\usepackage{enumitem}
\usepackage{rotfloat}
\usepackage{colortbl}
\usepackage{cleveref}

\usepackage{mathtools}

\newcommand{\R}{\mathbb{R}}

\newcommand{\W}{\mathbb{W}}

\renewcommand{\aa}{\mathbf{a}}

\newcommand{\xx}{\mathbf{x}}

\renewcommand{\AA}{\mathcal A}
\newcommand{\BB}{\mathcal B}

\newcommand{\FF}{\mathcal F}

\newcommand{\TT}{\mathcal T}

\begin{document}

\title{Adaptive Partitioning for Template Functions on Persistence Diagrams\\
\thanks{The work of ST and EM was supported in part by NSF grants DMS-1800446, CMMI-1800466, and CCF-1907591. FAK acknowledges the support of the National Science Foundation under grants CMMI-1759823 and DMS1759824. \newline\newline
\textcopyright 2019 IEEE. Personal use of this material is permitted.  Permission from IEEE must be obtained for all other uses, in any current or future media, including reprinting/republishing this material for advertising or promotional purposes, creating new collective works, for resale or redistribution to servers or lists, or reuse of any copyrighted component of this work in other works.
}
}
\author{\IEEEauthorblockN{Sarah Tymochko}
\IEEEauthorblockA{\textit{Dept. of Computational Mathematics,} \\ \textit{Science \& Engineering}  \\
\textit{Michigan State University}\\
East Lansing, MI \\
tymochko@egr.msu.edu}
\and
\IEEEauthorblockN{Elizabeth Munch}
\IEEEauthorblockA{\textit{Dept. of Computational Mathematics,} \\ \textit{Science \& Engineering} \\ \textit{and Dept. of Mathematics}\\
\textit{Michigan State University}\\
East Lansing, MI \\
muncheli@egr.msu.edu}
\and
\IEEEauthorblockN{Firas A. Khasawneh}
\IEEEauthorblockA{\textit{Dept. of Mechanical Engineering} \\
\textit{Michigan State University}\\
East Lansing, Michigan \\
khasawn3@egr.msu.edu}
}

\maketitle

\begin{abstract}
As the field of Topological Data Analysis continues to show success in theory and in applications, there has been increasing interest in using tools from this field with methods for machine learning.
Using persistent homology, specifically persistence diagrams, as inputs to machine learning techniques requires some mathematical creativity.
The space of persistence diagrams does not have the desirable properties for machine learning, thus methods such as kernel methods and vectorization methods have been developed.
One such featurization of persistence diagrams by Perea, Munch and Khasawneh in \cite{Perea2019} uses continuous, compactly supported functions, referred to as ``template functions,'' which results in a stable vector representation of the persistence diagram.
In this paper, we provide a method of adaptively partitioning persistence diagrams to improve these featurizations based on localized information in the diagrams.
Additionally, we provide a framework to adaptively select parameters required for the template functions in order to best utilize the partitioning method.
We present results for application to example data sets comparing classification results between template function featurizations with and without partitioning, in addition to other methods from the literature.

\end{abstract}


\section{Introduction}

The field of Topological Data Analysis (TDA) uses methods from algebraic topology to study the underlying shape of data.
Specifically, persistent homology is one tool that studies the homology of a changing space, resulting in a representation called a persistence diagram.
As persistent homology has shown success in many application fields, there has been significant interest in applying statistical and machine learning techniques to persistence diagrams directly.
However, the space of persistence diagrams lacks many of the desirable properties for machine learning tasks.
Specifically, the space of persistence diagrams is not a Banach space, and it does not have unique geodesics and thus has non-unique means.

Numerous methods have been developed to map persistence diagrams into a space more amenable for machine learning.
These methods include featurization methods, such as persistence images \cite{Adams2015} and persistence landscapes \cite{Bubenik2015}, along with many kernel functions \cite{Kusano2017, Reininghaus2015, Kwitt2015, Kusano2016, Carriere2017, Anirudh2016, Le2018PersistenceFK, Zhu2016}.
We will focus on one particular featurization using ``template functions'' \cite{Perea2019}.
A template function is any function on $\mathbb{R}^2$ that is continuous and compactly supported.
By evaluating a set of these template functions on a persistence diagram, we create a feature vector representation.
In this paper, we present a method of adaptively partitioning persistence diagrams in order to use the template function featurization on more localized regions of the persistence diagrams.
With this, we develop a method of adaptively modifying parameter choices for the functions to better fit the partitions.
This new adaptive method uses fewer features from the original template function featurization method.
In Sec.~\ref{ssec:templatefcns}, we present the original method for template functions and will describe our modifications of the method in Sec.~\ref{sec:methods}.
In Sec.~\ref{sec:applications}, we will present the results of our method on some example data sets compared to the results from \cite{Perea2019}.

\section{Background}
\label{sec:background}

Persistent homology is a method from TDA that studies how the homology changes as the space changes.
In this work we will briefly describe homology and how persistent homology can be applied to point cloud data.
While persistent homology can also be applied to data in the form of images or 3d voxel images, we will only use point cloud data in this paper.
We refer the interested reader to \cite{Edelsbrunner2010, Hatcher, Munch2017}.

\subsection{Homology and persistent homology}

Homology is a standard tool in algebraic topology to study topological structure in different dimensions.
In particular, given a space, $X$, homology computes a group for dimensions $k=0,1,2,\ldots$, denoted $H_k(X)$ that represent information about the structure in each dimension.
In particular, dimension 0 studies connected components, dimension 1 studies loops, dimension 2 studies voids, and higher dimensions study the higher dimensional analogues.

We will first introduce a few other concepts in order to define homology, specifically simplicial homology, more formally.
A simplicial complex, $\mathcal{K}$, is a space built from simplicies, where an $n$-simplex is the convex hull of $n+1$ affinely independent points.
The face of an $n$-simplex, $\sigma$, is defined to be the convex hull of a nonempty subset of the vertices of $\sigma$.
The simplical complex must satisfy the following requirements: (1) the intersection of any two simplices in $\mathcal{K}$ is also a simplex in $\mathcal{K}$ and (2) faces of a simplex in $\mathcal{K}$ are also simplices in $\mathcal{K}$.

For a given simplicial complex $\mathcal{K}$, let $\mathcal{K}_p$ be the set of all $p$-simplices, $p=0,1,2,\ldots$.
Then a $p$-chain, $c$, is defined to be a formal sum of $p$-simplices in $\mathcal{K}$,
\[
c = \sum_{\sigma_i \in \mathcal{K}_p} a_i \sigma_i,
\]
where coefficients $a_i \in \mathbb{Z}_2$.
Note that other fields can be used for coefficients, but we will focus on the simplified case of $\mathbb{Z}_2$ as that is typically what is used for persistent homology.
Since we can add and scale chains by a constant, the collection of $p$-chains, $C_p(\mathcal{K})$, called the chain group, forms a vector space.
The boundary map between chain groups is defined as the linear transformation
\[
\partial_p:C_p \to C_{p-1}
\]
which maps a $p$-simplex to the sum of its $(p-1)$-dimensional faces.
The chain complex is sequence of chain groups connected by the corresponding boundary maps,
\[
\cdots \xrightarrow{\partial_{p+2}} C_{p+1} \xrightarrow{\partial_{p+1}} C_{p} \xrightarrow{\partial_{p}} C_{p-1} \xrightarrow{\partial_{p-1}} \cdots.
\]

Within a chain group, we define two different kinds of $p$-chains that are needed to define the homology groups, cycles and boundaries.
A $p$-cycle is a $p$-chain, $c$, with $\partial_p(c) = 0$, meaning it has empty boundary.
The set of $p$-cycles is the kernel of the boundary map, $\text{ker}(\partial_p)$.
A $p$-boundary is a $p$-chain that is the boundary of a $p+1$-chain, i.e. $c_p = \partial c_{p+1}$ with $c_{p+1}\in C_{p+1}$.
The set of $p$-boundaries is the image of the boundary map, $\text{im}(\partial_p)$.
Note that the $p+1$-boundaries are a subgroup of the $p$-cycles.
Now, the $p$-th homology group is formally defined as
\[
H_p(\mathcal{K}) = \text{ker}(\partial_p) / \text{im}(\partial_{p+1}).
\]
Further, the $p$-th Betti number is defined to be the rank of the $p$-th homology group and is denoted $\beta_p$.

Persistent homology is a tool for studying the homology of a parameterized space.
We will define the Vietoris-Rips complex, which is one method of creating a simplicial complex based on a point cloud.
Given a point cloud, $X$, and a distance, $r$, the Vietoris-Rips complex is a simplicial complex where for every finite set of $n$ points with maximum pairwise distance at most $r$, the $n-1$ simplex formed by those points is added.

From the Vietoris-Rips complex, for every radius, $r_i$, we get a simplicial complex, $X_{r_i}$.
These complexes have the property that if $r_i<r_j$ then $X_{r_i} \subseteq X_{r_j}$.
Thus, for any increasing set of radii, $0<r_0<r_1<r_2< \cdots <r_n$, we get a nested sequence of simplicial complexes,
\begin{equation}
X_{r_0} \subseteq X_{r_1} \subseteq X_{r_2} \subseteq \cdots \subseteq X_{r_n}
\label{eqn:filtration}
\end{equation}
called a filtration.
Computing $k$ dimensional homology of each space in the filtration, the inclusions in (\ref{eqn:filtration}) induce linear maps between the homology groups,
\[
H_k(X_{r_0}) \to H_k(X_{r_1}) \to \cdots \to H_k(X_{r_n}).
\]
By studying these maps, we study how the homology of the space changes through the filtration.
In particular, we care about when features appear and disappear in this sequence.
We say a $k$-dimensional feature, $\gamma$ is ``born'' at radius $r_i$ if $\gamma \in H_k(x_{r_i})$ but $\gamma \not\in H_k(x_{r_{i-1}})$. A feature ``dies'' at radius $r_j$ if it merges with an older feature going from $X_{r_{j-1}}$ to $X_{r_j}$.

A persistence diagram is a way of representing the births and deaths of homology classes.
The persistence diagram is a scatter plot where a class that is born at $r_i$ and dies at $r_j$ is represented as the point $(r_i,r_j)$.
This is typically called the birth-death plane.
Since in a standard Vietoris-Rips filtration, a feature is always born before it dies, all persistence points are above the diagonal through the birth-death plane.
Another popular modification of a persistence diagram is to plot a class that is born at $r_i$ and dies at $r_j$ as the point $(r_i,r_j-r_i)$ where the quantity $r_j-r_i$ represents how long a feature lived, or its ``lifetime.''
This is called the birth-lifetime plane.
For simplicity of definitions to follow, we will be working in the birth-lifetime plane for the rest of the paper.

\subsection{Featurization Using Template Functions}
\label{ssec:templatefcns}

\begin{figure}[t]
  \centering
    \includegraphics[width=0.49\textwidth]{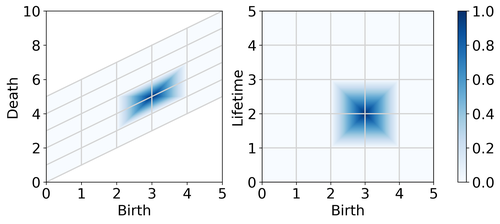}
  \caption{Example tent function, $g_{(3,2),1}$, drawn in the birth-death plane (left) and birth-lifetime plane (right) with $d=5$, $\delta=1$ and $\epsilon=0$. This plot is similar to \cite[Fig. 4]{Perea2019}.}
  \label{fig:tentheatmaps}
\end{figure}

\begin{figure}[t]
  \centering
    \includegraphics[width=0.45\textwidth]{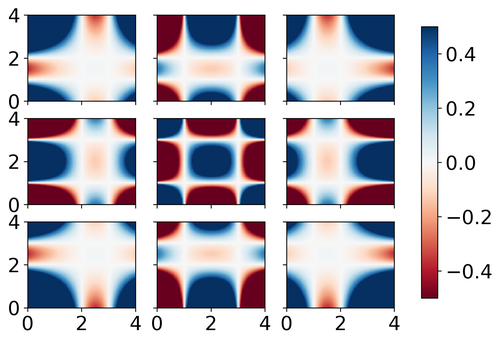}
\caption{Examples of interpolating polynomials for the meshes $\AA = \BB = \{1,2,3\}$. The plot drawn at $(i,j)$ shows the polynomial, $p_{i,j}$, where $p_{i,j} = 1$ and 0 on all other mesh points. This plot is from \cite[Fig. 5]{Perea2019}.}
\label{fig:polyheatmaps}
\end{figure}

As proposed in \cite{Perea2019}, we will consider featurizations of persistence diagrams based on two types of template functions, tent functions and interpolating polynomials.
We will briefly describe the method of featurization.

A \emph{template function} is defined as any function on $\mathbb{R}^2$ that that is continuous, and has compact support contained within the upper half plane, $\W:= \R \times \R_{>0}$ \footnote{Recall that we use birth-lifetime coordinates throughout the paper; otherwise, template functions could be equivalently defined to have support above the diagonal.}.
A template function $f:\W \to \R$ can be turned into a function on persistence diagrams as follows.
Given a diagram $D$, the function is evaluated on each point in the diagram, and then summed, giving
\begin{equation*}
    \nu_f (D) = \sum_{\xx \in D} f(\xx).
\end{equation*}

A collection of template functions, $\TT$, is called a \emph{template system} if the resulting functions on persistence diagrams, $\FF_\TT = \{ \nu_f: f \in \TT\}$ separate points.
That is, for every pair of diagrams, $D$ and $D'$, there exists a function $f \in \TT$ such that $\nu_f(D) \neq \nu_f(D')$.
As a true template system is infinite, vectorization is done by returning $(\nu_{f_1}(D), \cdots, \nu_{f_k}(D))$ for functions in some subset of the template system.
This is well justified since any function on persistence diagrams can be approximated by some finite subset of a template system; see \cite[Thm.~29]{Perea2019}.
In this paper, we will use two examples of template systems as given in \cite{Perea2019}: tent functions and interpolating polynomials.

Tent functions are an example of template functions that are meant to probe small regions of the persistence diagram.
Again, recall everything is defined in the birth-lifetime plane.
Given a point $\mathbf{a} = (a,b)$, and a radius $\delta \in \mathbb{R}_{>0}$ with $0 < \delta < b$, the tent function is defined to be
\[
g_{\mathbf{a},\delta}(x,y) = \left|1- \frac{1}{\delta} \max\{|x-a|,|y-b|\} \right|_{+},
\]
where $|\cdot|_{+}$ denotes the positive value of the function, and 0 otherwise.
This function is supported on the compact box $[a-\delta, a+\delta]\times[b-\delta, b+\delta]$, evaluates to $1$ at $\aa$, and decreases linearly to 0 on the boundary of the box.
Note that since the box must be compactly supported on persistence diagrams, the bottom edge of the box cannot lie on the $x$-axis.
Given a persistence diagram $D = \{\textbf{x} = (b_i,l_i)\}$, the tent function is the sum of the evaluation of this function on all points in the diagram,
\[
G_{\mathbf{a},\delta}(D) = \sum_{\mathbf{x}\in D}\ g_{\mathbf{a},\delta}(\mathbf{x}).
\]

The full template system consists of all tent functions $g_{\aa,\delta}$ which have compact support contained in $\mathbb{W}$.
However, in practice, we work with the subset of these tent functions
\begin{equation}
  \{ G_{(\delta i, \delta j + \epsilon), \delta} | 0 \leq i \leq d, 1 \leq j \leq d \}
\end{equation}
by choosing the grid size, $d$, and a vertical shift, $\epsilon$, to ensure $g$ is compactly supported inside $\W$.
This gives a $d\times (d+1)$ feature vector.
An example of a tent function is shown in Fig.~\ref{fig:tentheatmaps}.
In this figure, the grid represents the mesh on which tent functions can be centered.
We show a single tent function, centered at $(3,2)$ with $\delta=1$ and $\epsilon = 0$.

The second template system we work with are interpolating polynomials.
Unlike the localized tent functions, interpolating polynomials have support that fills out the space, however to satisfy the properties of template functions, they will be transformed to have compact support.
Given a mesh $\AA = \{a_i\}_{i=0}^m \subset \R$, the Lagrange polynomial $\ell_j^\AA(x)$ corresponding to $a_j$ is
\begin{equation*}
    \ell_j^\AA(x) =
    \prod_{i \neq j}
    \frac{x-a_i}{a_j-a_i}.
\end{equation*}
This has the property that $\ell_j^\AA(a_k)$ is 1 if $j=k$, and $0$ otherwise.
Then fixing meshes $\AA \subset \R$, $\BB \subset \R_{>0}$, and coordinates $i'$ and $j'$, the template function is
\begin{equation*}
    f(x,y) =  h(x,y)\cdot | \ell_{i'}^\AA(x) \ell_{j'}^\BB(y)|
\end{equation*}
where $h$ is a hill function forcing the resulting polynomial to have compact support inside a designated box.
In practice, the box for $h$ is a bounding box containing the mesh $\AA \times \BB$ where both meshes $\AA$ and $\BB$ are chosen to have $d$ elements; if this box further encloses all points in all diagrams, then its existence is implicit and need not be coded at all.
Examples of these interpolating polynomials are shown in Fig.~\ref{fig:polyheatmaps}.

\section{Methods}
\label{sec:methods}

\begin{figure}[t]
  \centerline{
    \includegraphics[width=0.24\textwidth]{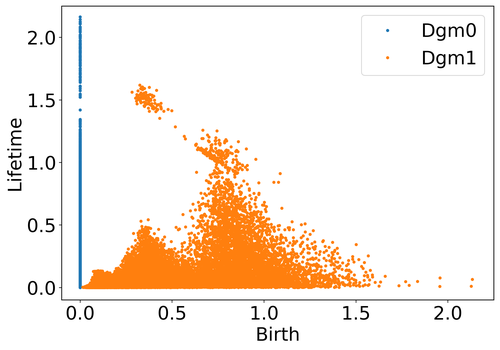}
    \includegraphics[width=0.24\textwidth]{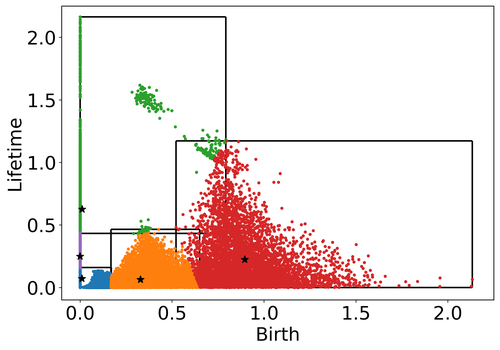}
  }
  \centerline{
    \includegraphics[width=0.24\textwidth]{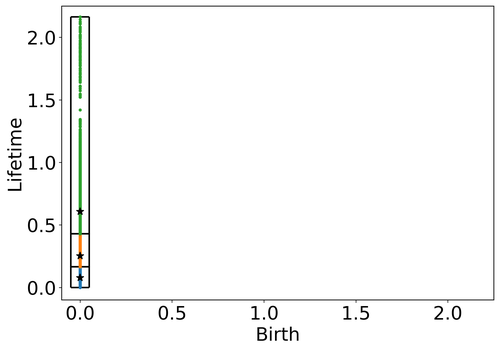}
    \includegraphics[width=0.24\textwidth]{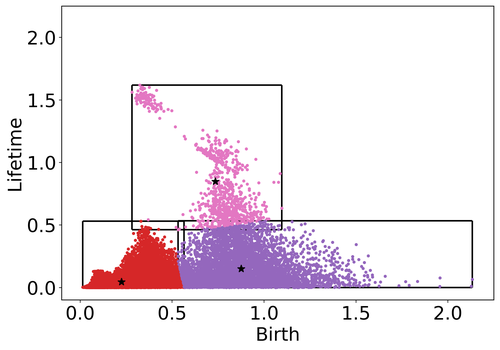}
  }
  \caption{The top left image is an example of a set of persistence diagrams from the manifold experiment explained in \ref{ssec:Manifolds} showing both the 0 and 1 dimensional diagrams in the birth-lifetime plane.
  The top right image is an example showing clustering on both 0 and 1 dimensional diagrams together, which we call ``combined partitioning,'' and creating 5 partitions.
  The bottom left and bottom right are examples showing 0 and 1 dimensional diagrams respectively, and clustering each dimension separately, which we call ``split partitioning,'' creating 3 partitions per dimension.
  In all except the top left image, the black stars represent centers of clusters from k-means clustering while the black boxes represent the partitions.}
  \label{fig:partitioning}
\end{figure}

The main contribution of this paper is to provide an adaptive method for choosing a subset of a template system based on K-means clustering \cite{hartigan1979algorithm}.
The method consists of two steps: first, cluster the points in all diagrams to find regions of interest, and second, construct localized template function systems based on these clusters.
We will ensure that the collection of these local template function systems covers all points in the diagrams.

To get the clusters, persistence diagrams in the training set are combined and input into the standard K-means clustering algorithm for a selected number of clusters $k$.
Clustering can be done in multiple ways.
For applications using several dimensions of diagrams, i.e.~0 and 1 dimensional diagrams, there are two possible options.
The first is combining all diagrams in the training set regardless of dimension;
the second is to combine only training persistence diagrams of like dimensions, and get a different set of clusters for each diagram dimension.
Figure~\ref{fig:partitioning} shows an example of a persistence diagram with both 0- and 1-dimensional persistence in the birth-lifetime plane along with examples showing the two different methods for generating clusters when using both 0 and 1 dimensional diagrams.
For simplicity, we will label results using the first option as ``combined partitioning'' while we will label results using the second option as ``split partitioning.''

Then, for each cluster, a covering box, which we call a \emph{partition}, is selected based on the bounding box of the points assigned to that particular cluster.
This results in one cover element per cluster; however, notice that the partitions themselves can overlap each other, and so points from the diagrams could land in the support of more than one partition.
For this reason, the clusters themselves are not particularly interesting, they are just used to select general regions where persistence points are located.
This method gives us a collection of partitions, each of which is a rectangular region in the birth-lifetime plane.
We then utilize a collection of template functions contained within each of these partitions in the same way we would have done in the original method when given only one bounding box.

\begin{figure}[t]
  \centerline{
    \includegraphics[width=0.12\textwidth]{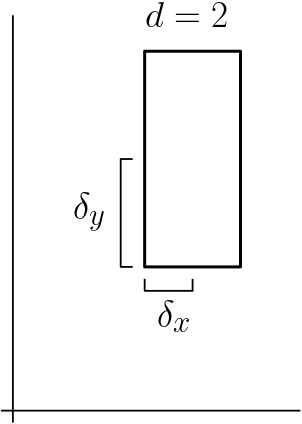}
    \includegraphics[width=0.12\textwidth]{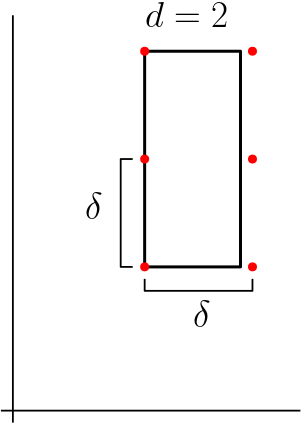}
    \includegraphics[width=0.15\textwidth]{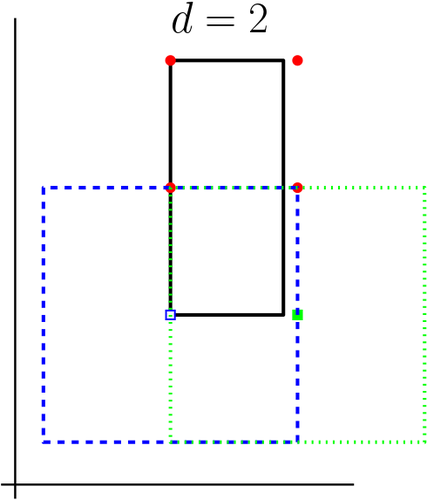}
  }
  \caption{Example of steps in adaptive parameter selection for a given partition, shown as the black rectangle. In this example, we are using tent functions with $d=2$. The leftmost image, we calculate $\delta_x$ and $\delta_y$ and choose $\delta$ to be the larger value. In the middle image, we select $d_y = 2$, calculate $d_x$ as explained in Sec.~\ref{sec:methods} which yields $d_x=1$, and apply a $(d_x+1)\times (d_y+1)$ grid (shown as the red points) where tent functions will be centered. In the rightmost image, for the tent centers that lie along the bottom of the partition (shown as a hollow blue square and solid green square), we check that the supports (shown as dashed and dotted boxes colored corresponding to their center) remain above the $x$-axis. Since they do, no further action is needed. An example of when the supports do cross the $x$-axis is shown in Fig.~\ref{fig:parameters2}.}
  \label{fig:parameters}
\end{figure}

We start by describing this process for the tent functions as defined in Sec.~\ref{ssec:templatefcns}, which have parameters $d, \delta,$ and $\epsilon$.
We develop a method of adaptively selecting $d$ and $\delta$ based on each partition, allowing for a more localized featurization.
In our modified version of the method, $d$ does not need to be the same in the $x$ and $y$ direction, thus we will write $d_x,d_y$ to specify the $d$ parameter in each.
Given a particular partition, $P = [x_{min}, x_{max}]\times [y_{min}, y_{max}]$, we first choose an initial value of parameter $d$.
From this, $\delta$ is calculated to be $\max\{\delta_x, \delta_y\}$ where $\delta_x = \frac{x_{max} - x_{min}}{d}$ and $\delta_y$ is defined similarly.
If $\delta_x > \delta_y$, then $d_x = d$ and $d_y = \lceil \frac{y_{max} - y_{min}}{\delta} \rceil$.
Similarly, if $\delta_x < \delta_y$ then $d_x = \lceil \frac{x_{max} - x_{min}}{\delta} \rceil$ and $d_y = d$.
Figure~\ref{fig:parameters} shows an example of this adaptive parameter selection process.
Note that by virtue of this notation, the support of the tent functions placed on the boundary of the partition extends outside the box.
This results in a grid of size $(d_x+1) \times(d_y+1)$ which reduces the number of features used per cover element yet ensures that based on the selected $d$ value that $\delta$ is selected appropriately to cover all points.

\begin{figure}[t]
  \centerline{
    \includegraphics[width=0.14\textwidth]{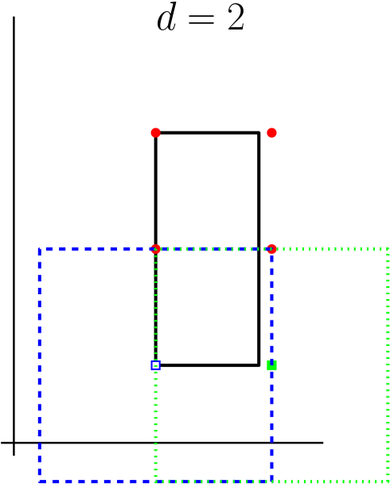}
    \includegraphics[width=0.14\textwidth]{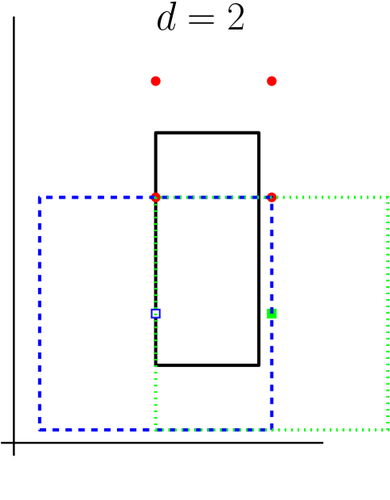}
    \includegraphics[width=0.14\textwidth]{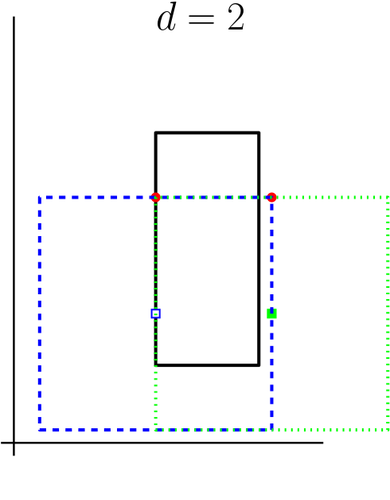}
  }
  \caption{Example of steps in adaptive parameter selection for a given partition, shown as the black rectangle, where the partition is close to the $x$-axis. In this example, we are using tent functions with $d=2$. The leftmost image, we apply the same process as in Fig.~\ref{fig:parameters} but the tent supports cross the $x$-axis. In the middle image, we shift up the grid where tent centers are placed so the tent support is at least a small $\epsilon>0$ above the $x$-axis. In the rightmost image, since the top two tent centers are more than $\delta/2$ outside the paritition, we remove them, decreasing $d_y$ by 1.}
  \label{fig:parameters2}
\end{figure}

Additional precautions are taken to ensure that the support of the tent functions did not cross the $x$-axis (or the diagonal in the birth-death plane).
Fix $\epsilon >0$, a parameter chosen by the user, then if after this parameter selection $y_{min} - \delta < 0$, then the grid of tent centers is shifted up to ensure the support of all tent functions is at least $\epsilon$ above the $x$-axis.
If in this shift, there are tent centers that are greater than $\delta/2$ above the partition boundary, then they are removed and $d_y$ is reduced by 1.
Figure~\ref{fig:parameters2} shows a visual example of this special case.

When using the interpolating polynomials, the same process as above is used to select $d_x$ and $d_y$.
We do not need a value of $\delta$ because the mesh is defined by a non-uniform Chebyshev mesh rather than using a regular grid like is done with the tent functions.
Note that for the tent functions, we allow $d_x$ and $d_y$ to be zero, resulting in a grid consisting of a single, row or column of tent centers, however for the interpolating polynomials we require at least a $2\times 2$ grid.

\section{Experiments}
\label{sec:applications}

\subsection{Code}

The python package \texttt{teaspoon}\footnote{https://github.com/lizliz/teaspoon} contains code for classification using tent functions and interpolating polynomials.
Classification was done using \texttt{RidgeClassifierCV} and \texttt{LogisticRegression} from the \texttt{sklearn} package \cite{scikit-learn}.

\subsection{Manifold Experiment}
\label{ssec:Manifolds}

\begin{table}[b]
  \centering
  \caption{Results of classification of manifold data as explained in Sec. \ref{ssec:Manifolds} using template functions with and without partitionin for different numbers of examples drawn from each type of manifold. Ridge regression is used for classification in both methods. Scores highlighted in blue give the best average score between the two methods.}
  \begin{tabular}{|c|cccc|}
    \hline
    & \multicolumn{4}{c|}{{Tents}} \\
    {Num} &
    \multicolumn{2}{c}{{No Partitioning}} &
    \multicolumn{2}{c|}{{Partitioning}} \\
    {Dgms} & Train & Test & Train & Test \\
    \hline
    10  & $99.8\% \pm 0.9$ & $96.5\% \pm 3.2$  & $100\% \pm 0.0$    & \cellcolor{cyan}$99.5\% \pm 1.5$  \\
    25  & $99.9\% \pm 0.3$ & $99.0\% \pm 1.0$  & $99.9\% \pm 0.3$   & \cellcolor{cyan}$99.6\% \pm 0.8$  \\
    50  & $99.9\% \pm 0.2$ & $99.9\% \pm 0.3$  & $100\% \pm 0.0$    & \cellcolor{cyan}$100\% \pm 0.0$   \\
    100 & $99.8\% \pm 0.1$ & $99.7\% \pm 0.4$  & $99.9\% \pm 0.1$   & \cellcolor{cyan}$99.8\% \pm 0.2$ \\
    200 & $99.5\% \pm 0.1$ & \cellcolor{cyan}$99.5\% \pm 0.3$      & $99.6\% \pm 0.1$ & $99.2\% \pm 0.3$  \\
    \hline
    & \multicolumn{4}{c|}{{Polynomials}} \\
    {Num} &
    \multicolumn{2}{c}{{No Partitioning}} &
    \multicolumn{2}{c|}{{Partitioning}} \\
    {Dgms} & Train & Test & Train & Test \\
    \hline
    10  &  $99.8\% \pm 0.9$ & $95.0\% \pm 3.9$ & $100\% \pm 0.0$  & \cellcolor{cyan}$97.5\% \pm 2.5$ \\
    25  &  $99.7\% \pm 0.5$ & $97.6\% \pm 1.5$ & $99.7\% \pm 0.5$ & \cellcolor{cyan}$99.4\% \pm 0.9$ \\
    50  &  $100\% \pm 0.0$  & $99.2\% \pm 0.9$ & $100\% \pm 0.1$  & \cellcolor{cyan}$99.5\% \pm 0.5$ \\
    100 &  $99.6\% \pm 0.2$ & $99.3\% \pm 0.5$ & $99.7\% \pm 0.2$ & \cellcolor{cyan}$99.6\% \pm 0.5$ \\
    200 &  $99.2\% \pm 0.2$ & $98.9\% \pm 0.5$ & $99.5\% \pm 0.2$ & \cellcolor{cyan}$99.4\% \pm 0.3$ \\
    \hline
  \end{tabular}
  \label{tab:ManifoldRidge}
\end{table}

Replicating an experiment from \cite{Adams2015,Perea2019}, we generated collections of point clouds drawn from different manifolds.
Each point cloud consists of 200 points drawn from the following manifolds:

\begin{itemize}
    \item \textbf{Annulus:} points drawn uniformly from an annulus with inner radius 1 and outer radius 2.

    \item \textbf{Torus:} points drawn uniformly from a torus created from a rotating circle of radius 1 in the $xz$-plane centered at $(2,0)$ around the $z$-axis.

    \item \textbf{Sphere:} points drawn from a sphere in $\mathbb{R}^3$ of radius 1. Uniform noise in $[-0.05,0.05]$ was added to the radius.

    \item \textbf{Cube:} points drawn uniformly from $[0,1]^2\subset \mathbb{R}^2$.

    \item \textbf{3 Clusters:} points drawn from one of three different normal distributions with means $(0,0), (0,2), (2,0)$, each with standard deviation of 0.05.

    \item \textbf{3 Clusters of 3 Clusters:} points drawn from normal distributions centered at (0,0), (0,1.5), (1.5,0), (0,4), (1,3), (1,5), (3,4), (3,5.5), (4.5,5) each with standard deviation 0.05.
\end{itemize}

These point clouds can be generated using the function
\texttt{MakeData.PointCloud.testSetManifolds} in \texttt{teaspoon}.

\begin{figure}[tb]
  \centering
  \includegraphics[width=0.45\textwidth]{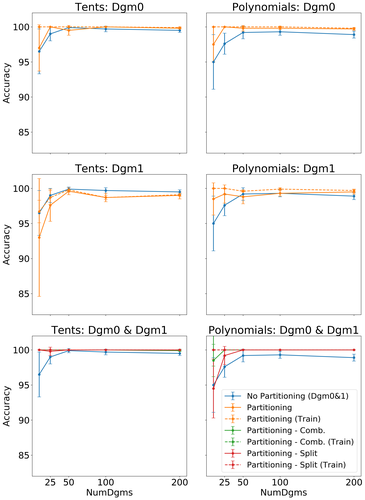}
  \caption{Results of classification of manifold data as explained in Sec. \ref{ssec:Manifolds} using template functions with and without partitioning. For partitioning methods, classification is done using logistic regression. Note that in all plots, the results without partitioning represent the accuracy using 0- and 1-dimensional diagrams. Thus the same accuracy is shown in each plot in a column. }
  \label{fig:ManifoldResults}
\end{figure}

\begin{figure}[tb]
  \centering
  \includegraphics[width=0.45\textwidth]{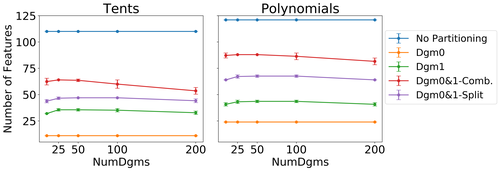}
  \caption{Average number of features used for the manifold experiment using template functions with and without partitioning. These correspond to the classification accuracies shown in Fig.~\ref{fig:ManifoldResults}.}
  \label{fig:ManifoldFeatures}
\end{figure}

For our method we tested a variety of parameters and options. For all experiments, we reserve $33\%$ of the data for testing while the remaining was used for training.
All the classification results are averaged over 10 runs of the experiment to control for outliers.

For comparison, Table~\ref{tab:ManifoldRidge} shows the results from \cite{Perea2019} using 0- and 1-dimensional diagrams with ridge regression for classification along with our accuracies using our partitioning method where partitions are selected based on both diagram dimensions simultaneously, referred to throughout the paper as ``combined partitioning.''
For the results from \cite{Perea2019}, the authors used tent parameters of $d=10$, $\epsilon$ to be half the minimum lifetime over all training set diagrams, and $\delta$ to be chosen to ensure the bounding box covered the training diagrams.
For our results, we used 3 clusters resulting in 3 partitions and for both template functions we set a starting value of $d=3$ and set $\epsilon$ to be machine precision, while the additional parameters are selected as described in Sec.~\ref{sec:methods}.
Note that in all cases except one, our partitioning method has a higher testing accuracy.

Figure~\ref{fig:ManifoldResults} shows the results of using our partitioning method on only 0- or 1-dimensional diagrams, on both dimensions using combined partitioning, and on both dimensions using split partitioning.
Here we still used 3 clusters resulting in 3 partitions, while starting with a value of $d=3$ and for both template functions.
For these experiments we used logistic regression for classification.
It is important to note that in \cite{Perea2019}, only accuracies using both dimensions were reported so the accuracies for no partitioning in all plots are from using both dimensions of diagrams with classification done using ridge regression.
This means that those results are the same across all plots for a given template function.

Using both 0- and 1-dimensional diagrams with either split or combined parititoning, using both tent functions and interpolating polynomials we get above 99\% accuracy for all except a couple cases.
Using only 0-dimensional or only 1-dimensional diagrams, we still get very good accuracy, but interestingly using 0-dimensional diagrams, the accuracies seem slightly better.
Using only 0-dimensional diagrams, we almost always outperform the template functions without partitioning using both 0- and 1-dimensional diagrams.
However, using tent functions with 1-dimensional diagrams, our method underperforms.
Additionally, Fig.~\ref{fig:ManifoldFeatures} shows the average number of features used for these experiments.
The number of features used is dependent on which diagrams are being used.
For example, using only 0-dimensional diagrams, we need very few features as all points in the diagrams fall on the $y$-axis for this experiment.
Using both 0- and 1-dimensional diagrams will require more features as we need to cover more of the diagram.
However in all cases, we are using significantly less features and still achieving comparable or higher accuracies.

\subsection{Shape Data}
\label{ssec:ShapeData}

\begin{figure}[tb]
  \centering
  \includegraphics[width=0.45\textwidth]{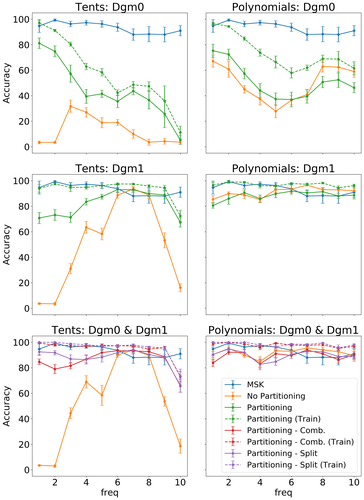}
  \caption{Results of classification of shape data as explained in Sec. \ref{ssec:ShapeData} using template functions with and without partitioning. In both columns, MSK gives the original results from \cite{Reininghaus2015}. Ridge regression is used for classification.}
  \label{fig:ShapeRidge}
\end{figure}

\begin{figure}[tb]
  \centering
  \includegraphics[width=0.49\textwidth]{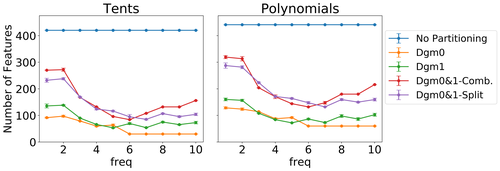}
  \caption{Average number of features used for shape data experiment using our partitioning method. These correspond to the classification accuracies shown in Fig.~\ref{fig:ShapeRidge}.}
  \label{fig:ShapeRidge-Features}
\end{figure}

As in \cite{Perea2019}, we compared our results to the kernel method developed in \cite{Reininghaus2015} on the synthetic SHREC 2014 data set \cite{SHREC}.
This data set consists of 3D meshes of fifteen humans (five males, five females and five children) in 20 different poses.
In \cite{Reininghaus2015}, the authors define a function on each mesh using the heat kernel signature for 10 different parameters and compute 0- and 1-dimensional diagrams.
Using the 300 pairs of persistence diagrams for each 10 parameter values, we predict which of the 15 humans is represented in each mesh.

Figure~\ref{fig:ShapeRidge} show the results of these experiments using our partitioning method with tent and interpolating polynomial functions as well as the results reported in \cite{Perea2019} (labeled as ``No Partitioning'') and \cite{Reininghaus2015} (labeled as ``MSK'').
For clarity, tables showing these results are also located in the appendix in Tables~\ref{tab:ShapeResults}.
For all experiments with partitioning, we use $d=5$ and 5 clusters for partitioning.
For the experiments without partitioning, the authors use $d=20$ for both types of template functions.
All accuracies with and without partitioning are averaged over 10 runs.

For three of the ten parameter values, using tent functions, our method achieves the highest accuracy.
Additionally, the confidence intervals intersect the highest accuracy for three additional parameters using tent functions.
Comparing tent functions with and without partitioning, it is clear that partitioning drastically improves most of the accuracies.
For example, using 0- or 1-dimensional diagrams, the top left and middle left plot in Fig.~\ref{fig:ShapeRidge}, the green line representing our testing accuracy is almost always higher than the orange line representing the testing accuracy without partitioning.

Using interpolating polynomials, our method does not surpass the kernel method or the template functions without partitioning to achieve the highest accuracy, however the confidence intervals intersect the highest accuracy for seven of the ten parameters.
Comparing our results to featurization using interpolating polynomial functions without partitioning, our results are fairly comperable; for some parameter values we achieve slightly higher accuracies, while for others we achieve slightly lower accuracies.
Without partitioning, the interpolating polynomials are not localized and may be picking up more global structure in the diagrams that is missed using partitioning, which could explain this lack of drastic improvement.

Figure~\ref{fig:ShapeRidge-Features} shows the average number of features used for these experiments.
It is clear that we are using far less than the 420 and 441 features used in \cite{Perea2019} with tent and polynomial functions respectively, yet we still achieve good accuracies, particularly using tent functions.
For example, for the parameters where tent functions with partitioning achieve the highest accuracy, we use less than 150 features.

\section{Discussion}

In this paper, we have presented a method of adaptively partitioning persistence diagrams for featurization using template functions.
This allows for significant flexibility in the featurization for the user.
Our methods are applicable to tent function and interpolating polynomial template systems, but additionally to any functions fitting into the template framework from \cite{Perea2019}.
We also provide two different methods of featurizing using several dimensions of diagrams, either partitioning all dimensions together, or separately.

For further modification of this method, the $k$-means clustering algorithm could be changed to any clustering technique as chosen by the user.
$K$-means clustering, and other clustering algorithms can also take weights into account, allowing a user to weight certain points in the diagram more heavily than others.
A standard example is weighting points based on the lifetime, as features with longer lifetimes are typically interpreted as the most significant, while features with shorter lifetimes are often considered noise.

We have shown that our method works well on standard examples in comparison to \cite{Perea2019} and other featurization methods available.
In particular, our method improves on the results of template functions without partitioning for the manifold experiment, achieving above 99\% accuracy in almost all cases.
For the SHREC data set, our method improves drastically upon results without partitioning when used with tent functions however is not quite as successful using interpolating polynomials.
In \cite{Perea2019} the authors point out the instinct to emphasize localization for these persistence diagram featurizations.
For tent functions, which are defined already to be very localized, our method allows for even closer analysis of regions with high density of points, ignoring more sparse regions.
However, the interpolating polynomials are more global, where each function in the template system has support over the entire region of points in the diagram when not using partitions.
In \cite{Perea2019}, this seems to be to their advantage as the interpolating polynomials significantly outperform tent functions on the SHREC data set.
The fact that these functions worked so well in \cite{Perea2019} is surprising, but it may indicate that trying to localize these functions by restricting their support to the selected partitions may cause global information to be overlooked.

In general, partitioning the diagrams allows for a more localized featurization of the diagrams, as is desirable in many applications.
Our method allows for a new adaptation of an existing featurization method which gives the user significant freedom to test parameters, existing examples of template functions, and any other functions that fit the template system criteria.

\bibliographystyle{IEEEtran}
\bibliography{../TemplatePartitioning}

\appendix
\label{appendix}

This appendix gives a table with the accuracies for classification of the shape data using template functions with and without partitioning.
The data shown in Table~\ref{tab:ShapeResults} is represented graphically in Fig.~\ref{fig:ShapeRidge}.

\begin{table*}[ht]
  \caption{Results of classification of shape data as explained in Sec. \ref{ssec:ShapeData} using tent functions and polynomial functions, with and without partitioning. Ridge regression is used for classification. The MSK column gives the original results from \cite{Reininghaus2015}. Scores highlighted in blue give the best average score across all testing columns; scores highlighted in orange have overlapping intervals of standard deviation with the best score.}
  \begin{center}
  \begin{tabular}{|c|c|cc|cc|cc|cc|}
    \hline
    &
    & \multicolumn{8}{c|}{{Tents - Partitioning}} \\
    & &
    \multicolumn{2}{c}{{Dgm0}}
    & \multicolumn{2}{c}{{Dgm1}}
    & \multicolumn{2}{c}{{Dgm0 \& Dgm1}}
    & \multicolumn{2}{c|}{{Dgm0 \& Dgm1}} \\
    & & \multicolumn{2}{c}{{}}
    & \multicolumn{2}{c}{{}}
    & \multicolumn{2}{c}{{(Combined Partitioning)}}
    & \multicolumn{2}{c|}{{(Split Partitioning)}} \\
    Freq & MSK & Train & Test & Train & Test & Train & Test & Train & Test \\
    \hline
    1 & \cellcolor{cyan}94.7 $\pm$ 5.1
    & 97.1 $\pm$ 1.8 & 81.2 $\pm$ 4.2
    & 93.9 $\pm$ 2.9 & 70.8 $\pm$ 4.4
    & 99.4 $\pm$ 0.4 & 84.7 $\pm$ 2.2
    & 100.0 $\pm$ 0.0 & \cellcolor{orange!25}92.5 $\pm$ 2.0 \\

    2 & \cellcolor{cyan}99.3 $\pm$ 0.9
    & 91.2 $\pm$ 0.9 & 74.8 $\pm$ 4.3
    & 97.5 $\pm$ 0.8 & 73.2 $\pm$ 4.3
    & 97.7 $\pm$ 0.7 & 79.0 $\pm$ 3.6
    & 100.0 $\pm$ 0.0 & 91.8 $\pm$ 1.8 \\

    3 & \cellcolor{cyan}96.3 $\pm$ 2.2
    & 80.4 $\pm$ 1.7 & 57.3 $\pm$ 6.7
    & 94.9 $\pm$ 3.4 & 71.3 $\pm$ 4.0
    & 97.4 $\pm$ 0.8 & 81.6 $\pm$ 2.4
    & 98.8 $\pm$ 0.5 & 86.9 $\pm$ 3.3 \\

    4 & \cellcolor{cyan}97.3 $\pm$ 1.9
    & 62.9 $\pm$ 2.5 & 39.5 $\pm$ 5.5
    & 94.8 $\pm$ 1.4 & 83.5 $\pm$ 2.6
    & 96.6 $\pm$ 1.0 & 86.8 $\pm$ 3.0
    & 98.1 $\pm$ 0.8 & 86.5 $\pm$ 3.7 \\

    5 & \cellcolor{cyan}96.3 $\pm$ 2.5
    & 58.4 $\pm$ 2.9 & 41.5 $\pm$ 2.8
    & 96.2 $\pm$ 1.7 & 87.5 $\pm$ 1.9
    & 97.6 $\pm$ 1.1 & \cellcolor{orange!25}91.9 $\pm$ 2.3
    & 96.9 $\pm$ 1.7 & 88.3 $\pm$ 3.8 \\

    6 & \cellcolor{cyan}93.7 $\pm$ 3.2
    & 42.3 $\pm$ 2.3 & 35.6 $\pm$ 5.0
    & 97.5 $\pm$ 0.9 & \cellcolor{orange!25}93.1 $\pm$ 1.8
    & 97.3 $\pm$ 0.9 & \cellcolor{orange!25}93.4 $\pm$ 2.6
    & 97.3 $\pm$ 1.0 & \cellcolor{orange!25}91.9 $\pm$ 2.5 \\

    7 & \cellcolor{orange!25}88.0 $\pm$ 4.5
    & 48.6 $\pm$ 2.6 & 43.7 $\pm$ 3.0
    & 97.4 $\pm$ 0.7 & \cellcolor{orange!25}92.9 $\pm$ 2.0
    & 97.1 $\pm$ 0.8 & \cellcolor{orange!25}93.3 $\pm$ 2.4
    & 97.9 $\pm$ 0.9 & \cellcolor{cyan}94.2 $\pm$ 2.3 \\

    8 & \cellcolor{orange!25}88.3 $\pm$ 6.0
    & 47.4 $\pm$ 3.6 & 36.6 $\pm$ 7.0
    & 95.9 $\pm$ 1.0 & \cellcolor{orange!25}89.9 $\pm$ 2.3
    & 94.6 $\pm$ 1.8 & \cellcolor{cyan}92.6 $\pm$ 2.0
    & 96.2 $\pm$ 0.8 & \cellcolor{orange!25}90.4 $\pm$ 3.0 \\

    9 & \cellcolor{orange!25}88.0 $\pm$ 5.8
    & 35.9 $\pm$ 11.8 & 25.8 $\pm$ 10.8
    & 94.5 $\pm$ 1.9 & \cellcolor{cyan}88.9 $\pm$ 3.0
    & 95.8 $\pm$ 1.0 & \cellcolor{orange!25}88.7 $\pm$ 1.9
    & 95.6 $\pm$ 1.4 & \cellcolor{orange!25}88.1 $\pm$ 2.5 \\

    10 & \cellcolor{cyan}91.0 $\pm$ 4.0
    & 11.3 $\pm$ 4.8 & 5.2 $\pm$ 3.4
    & 72.4 $\pm$ 4.4 & 67.3 $\pm$ 3.6
    & 73.1 $\pm$ 3.8 & 65.8 $\pm$ 5.0
    & 73.6 $\pm$ 5.4 & 66.3 $\pm$ 5.4 \\
    \hline
    \end{tabular}
    \end{center}

    \begin{center}
    \begin{tabular}{|c|cc|cc|cc|}
      \hline
      & %
      \multicolumn{6}{c|}{{Tents - No Partitioning}} \\
       & %
      \multicolumn{2}{c}{{Dgm0}}
      & \multicolumn{2}{c}{{Dgm1}}
      & \multicolumn{2}{c|}{{Dgm0 \& Dgm1}}\\
      Freq %
      & Train & Test & Train & Test & Train & Test \\
      \hline
      1 %
      & 8.3 $\pm$ 0.5 & 3.4 $\pm$ 1.1
      & 8.1 $\pm$ 0.2 & 3.7 $\pm$ 0.5
      & 8.2 $\pm$ 0.3 & 3.5 $\pm$ 0.5 \\
      2 %
      & 8.3 $\pm$ 0.3 & 3.4 $\pm$ 0.7
      & 8.2 $\pm$ 0.5 & 3.5 $\pm$ 1.1
      & 8.6 $\pm$ 0.4 & 3.0 $\pm$ 1.0 \\
      3 %
      & 66.5 $\pm$ 2.7 & 31.8 $\pm$ 4.8
      & 50.6 $\pm$ 2.1 & 31.1 $\pm$ 4.0
      & 80.5 $\pm$ 1.3 & 44.4 $\pm$ 4.3 \\
      4 %
      & 46.2 $\pm$ 2.5 & 27.0 $\pm$ 3.8
      & 83.1 $\pm$ 1.6 & 63.5 $\pm$ 4.6
      & 89.1 $\pm$ 1.5 & 69.0 $\pm$ 4.9 \\
      5 %
      & 28.5 $\pm$ 1.4 & 18.9 $\pm$ 4.0
      & 75.2 $\pm$ 2.6 & 58.3 $\pm$ 4.6
      & 76.8 $\pm$ 2.7 & 58.4 $\pm$ 7.9 \\
      6 %
      & 25.4 $\pm$ 1.8 & 19.0 $\pm$ 2.4
      & 96.5 $\pm$ 1.1 & \cellcolor{orange!25}88.7 $\pm$ 2.4
      & 96.8 $\pm$ 0.7 & \cellcolor{orange!25}89.9 $\pm$ 1.7 \\
      7 %
      & 19.4 $\pm$ 2.6 & 10.0 $\pm$ 3.4
      & 98.2 $\pm$ 0.5 & \cellcolor{orange!25}93.6 $\pm$ 1.9
      & 98.3 $\pm$ 0.6 & \cellcolor{orange!25}94.1 $\pm$ 2.5 \\
      8 %
      & 10.8 $\pm$ 2.6 & 3.6 $\pm$ 2.4
      & 91.9 $\pm$ 0.9 & \cellcolor{orange!25}88.8 $\pm$ 2.7
      & 91.9 $\pm$ 1.2 & \cellcolor{orange!25}89.7 $\pm$ 3.3  \\
      9 %
      & 10.6 $\pm$ 2.7 & 4.3 $\pm$ 2.2
      & 63.8 $\pm$ 2.7 & 53.3 $\pm$ 5.9
      & 64.9 $\pm$ 2.3 & 53.7 $\pm$ 3.8 \\
      10 %
      & 9.2 $\pm$ 2.3 & 3.6 $\pm$ 1.7
      & 27.0 $\pm$ 3.9 & 16.2 $\pm$ 3.2
      & 27.3 $\pm$ 3.4 & 18.6 $\pm$ 5.6 \\
      \hline
    \end{tabular}
    \end{center}

    \vspace{0.3cm}

  \begin{center}
    \begin{tabular}{|c|c|cc|cc|cc|cc|}
      \hline
      &
      & \multicolumn{8}{c|}{{Polynomials - Partitioning}} \\
      & &
      \multicolumn{2}{c}{{Dgm0}}
      & \multicolumn{2}{c}{{Dgm1}}
      & \multicolumn{2}{c}{{Dgm0 \& Dgm1}}
      & \multicolumn{2}{c|}{{Dgm0 \& Dgm1}} \\
      & & \multicolumn{2}{c}{{}}
      & \multicolumn{2}{c}{{}}
      & \multicolumn{2}{c}{{(Combined Partitioning)}}
      & \multicolumn{2}{c|}{{(Split Partitioning)}} \\
      Freq & MSK & Train & Test & Train & Test & Train & Test & Train & Test \\
      \hline
      1 & \cellcolor{cyan}94.7 $\pm$ 5.1
      & 96.6 $\pm$ 1.0 & 75.3 $\pm$ 5.2
      & 96.5 $\pm$ 2.2 & 80.6 $\pm$ 2.1
      & 98.8 $\pm$ 1.2 & 83.8 $\pm$ 2.8
      & 100.0 $\pm$ 0.0 & \cellcolor{orange!25}90.4 $\pm$ 2.4 \\

      2 & \cellcolor{cyan}99.3 $\pm$ 0.9
      & 94.4 $\pm$ 0.6 & 72.4 $\pm$ 4.2
      & 99.2 $\pm$ 1.2 & 86.0 $\pm$ 4.4
      & 99.9 $\pm$ 0.2 & 92.2 $\pm$ 1.9
      & 100.0 $\pm$ 0.0 & 94.8 $\pm$ 1.4 \\

      3 & \cellcolor{cyan}96.3 $\pm$ 2.2
      & 84.8 $\pm$ 1.4 & 57.4 $\pm$ 4.2
      & 98.8 $\pm$ 0.7 & 90.6 $\pm$ 2.5
      & 98.7 $\pm$ 0.9 & \cellcolor{orange!25}91.9 $\pm$ 2.7
      & 100.0 $\pm$ 0.1 & \cellcolor{orange!25}92.0 $\pm$ 2.7 \\

      4 & \cellcolor{cyan}97.3 $\pm$ 1.9
      & 73.8 $\pm$ 2.8 & 44.1 $\pm$ 5.5
      & 96.2 $\pm$ 1.9 & 86.0 $\pm$ 2.8
      & 96.0 $\pm$ 1.9 & 83.1 $\pm$ 3.2
      & 97.9 $\pm$ 0.9 & 82.7 $\pm$ 4.3 \\

      5 & \cellcolor{cyan}96.3 $\pm$ 2.5
      & 66.4 $\pm$ 4.7 & 37.4 $\pm$ 4.8
      & 95.8 $\pm$ 2.0 & 89.8 $\pm$ 3.1
      & 99.3 $\pm$ 0.5 & \cellcolor{orange!25}91.0 $\pm$ 3.1
      & 96.4 $\pm$ 1.2 & 84.8 $\pm$ 3.0 \\

      6 & \cellcolor{cyan}93.7 $\pm$ 3.2
      & 57.9 $\pm$ 4.0 & 37.0 $\pm$ 5.8
      & 97.9 $\pm$ 0.7 & \cellcolor{orange!25}91.8 $\pm$ 2.3
      & 97.8 $\pm$ 0.8 & \cellcolor{orange!25}89.6 $\pm$ 2.3
      & 98.6 $\pm$ 0.4 & \cellcolor{orange!25}90.5 $\pm$ 3.2 \\

      7 & 88.0 $\pm$ 4.5
      & 61.9 $\pm$ 2.2 & 39.5 $\pm$ 5.4
      & 97.2 $\pm$ 1.0 & 90.5 $\pm$ 2.3
      & 97.5 $\pm$ 0.8 & 93.9 $\pm$ 1.6
      & 98.3 $\pm$ 0.5 & 92.9 $\pm$ 1.7 \\

      8 & \cellcolor{orange!25}88.3 $\pm$ 6.0
      & 69.0 $\pm$ 2.9 & 50.8 $\pm$ 4.5
      & 98.2 $\pm$ 0.7 & \cellcolor{orange!25}91.4 $\pm$ 2.0
      & 98.1 $\pm$ 0.9 & \cellcolor{orange!25}91.0 $\pm$ 3.2
      & 99.4 $\pm$ 0.5 & \cellcolor{orange!25}92.6 $\pm$ 1.8 \\

      9 & \cellcolor{orange!25}88.0 $\pm$ 5.8
      & 68.7 $\pm$ 6.1 & 52.5 $\pm$ 6.1
      & 94.5 $\pm$ 1.5 & 88.1 $\pm$ 2.0
      & 95.2 $\pm$ 1.3 & 86.3 $\pm$ 2.7
      & 95.7 $\pm$ 2.2 & \cellcolor{orange!25}88.9 $\pm$ 3.5 \\

      10 & \cellcolor{orange!25}91.0 $\pm$ 4.0
      & 61.5 $\pm$ 5.1 & 46.3 $\pm$ 4.0
      & 96.2 $\pm$ 1.1 & 88.7 $\pm$ 3.0
      & 96.8 $\pm$ 1.2 & \cellcolor{orange!25}90.1 $\pm$ 2.5
      & 97.9 $\pm$ 1.0 & 88.8 $\pm$ 2.6 \\
      \hline
    \end{tabular}
  \end{center}
  \begin{center}
    \begin{tabular}{|c|cc|cc|cc|}
      \hline
      & %
      \multicolumn{6}{c|}{{Polynomials - No Partitioning}} \\
      & %
      \multicolumn{2}{c}{{Dgm0}}
      & \multicolumn{2}{c}{{Dgm1}}
      & \multicolumn{2}{c|}{{Dgm0 \& Dgm1}}\\
      Freq %
      & Train & Test & Train & Test & Train & Test \\
      \hline
      1 %
      & 94.3 $\pm$ 0.5 & 67.1 $\pm$ 4.7
      & 99.1 $\pm$ 0.3 & 85.4 $\pm$ 3.0
      & 99.8 $\pm$ 0.3 & \cellcolor{orange!25}90.4 $\pm$ 5.3 \\

      2 %
      & 92.1 $\pm$ 1.4 & 60.8 $\pm$ 6.3
      & 99.9 $\pm$ 0.3 & 89.9 $\pm$ 1.5
      & 100.0 $\pm$ 0.0 & 95.1 $\pm$ 2.4 \\

      3 %
      & 83.4 $\pm$ 2.4 & 45.1 $\pm$ 2.9
      & 99.6 $\pm$ 0.5 & 88.9 $\pm$ 3.0
      & 99.7 $\pm$ 0.5 & 90.0 $\pm$ 2.0 \\

      4 %
      & 74.7 $\pm$ 2.0 & 37.4 $\pm$ 4.7
      & 99.1 $\pm$ 0.7 & 85.2 $\pm$ 2.5
      & 98.6 $\pm$ 0.9 & 84.8 $\pm$ 3.9  \\

      5 %
      & 65.3 $\pm$ 2.9 & 27.8 $\pm$ 5.0
      & 99.2 $\pm$ 0.7 & \cellcolor{orange!25}93.0 $\pm$ 2.2
      & 99.7 $\pm$ 0.4 & \cellcolor{orange!25}93.3 $\pm$ 2.2  \\

      6 %
      & 67.2 $\pm$ 2.5 & 36.5 $\pm$ 3.6
      & 99.2 $\pm$ 0.5 & \cellcolor{orange!25}93.4 $\pm$ 2.8
      & 98.8 $\pm$ 0.5 & \cellcolor{orange!25}92.9 $\pm$ 1.8 \\

      7 %
      & 71.5 $\pm$ 2.8 & 40.9 $\pm$ 4.1
      & 98.3 $\pm$ 0.7 & \cellcolor{cyan}96.6 $\pm$ 0.7
      & 99.0 $\pm$ 0.4 & \cellcolor{orange!25}95.6 $\pm$ 1.4  \\

      8 %
      & 84.2 $\pm$ 3.3 & 63.0 $\pm$ 4.5
      & 99.0 $\pm$ 0.5 & \cellcolor{orange!25}93.0 $\pm$ 1.8
      & 99.6 $\pm$ 0.4 & \cellcolor{cyan}94.0 $\pm$ 2.2  \\

      9 %
      & 83.5 $\pm$ 2.7 & 62.4 $\pm$ 5.0
      & 98.4 $\pm$ 1.2 & \cellcolor{cyan}92.9 $\pm$ 1.5
      & 98.5 $\pm$ 1.3 & \cellcolor{orange!25}92.6 $\pm$ 2.1 \\

      10 %
      & 79.8 $\pm$ 2.7 & 59.0 $\pm$ 4.6
      & 96.9 $\pm$ 0.6 & \cellcolor{cyan}92.1 $\pm$ 1.7
      & 97.7 $\pm$ 1.1 & \cellcolor{orange!25}89.5 $\pm$ 4.6 \\
      \hline
    \end{tabular}
  \end{center}
  \label{tab:ShapeResults}
\end{table*}

\end{document}